\documentclass[prb,preprint,showpacs,preprintnumbers,amsmath,amssymb]{revtex4}

\usepackage{graphicx}
\usepackage{dcolumn}

\begin{document}
\bibliographystyle{revtex}
\draft


\title{Pseudogap formation and superconductivity of Bi$_{1.74}$Pb$_{0.38}$Sr$_{1.88}$CuO$_{6+\delta}$ 
by studies of out-of-plane resistivity in magnetic fields}



\author{K. Kudo}
\email{kudo@imr.tohoku.ac.jp}
\author{Y. Miyoshi}
\author{T. Sasaki}
\author{N. Kobayashi}
\affiliation{Institute for Materials Research, Tohoku University, Katahira 2-1-1, Aoba-ku, Sendai 980-8577, Japan}


\date{\today}

\begin{abstract}
We report the out-of-plane resistivity in the systematically oxygen-controlled single crystals Bi$_{1.74}$Pb$_{0.38}$Sr$_{1.88}$CuO$_{6+\delta}$ in magnetic fields parallel to the $c$-axis. 
Two characteristic temperatures $T^*$ and $T^{**}$ are found at which a semiconductor-like upturn starts to develop and the magnetoresistance changes its sign from positive to negative at lower temperature, respectively. 
The obtained phase diagram strongly suggests that the pseudogap temperature correlating to the superconducting gap is not $T^{*}$ but $T^{**}$. 
The gap opened at $T^*$ may relate to another phenomenon with the different energy scale such as the antiferromagnetic excitation due to the exchange interaction between Cu$^{2+}$ spins.
\end{abstract}

\maketitle

The pseudogap phenomena have attracted considerable attention, because it is believed to be a key mechanism to understand the high temperature superconductivity.\cite{rf:Timusk} 
There exist various aspects to understand the pseudogap, which are classified into two groups, such as the precursor of the superconductivity\cite{rf:Randeria,rf:Emery,rf:Kao,rf:Iyenger} and the competing energy gap or hidden order parameter to the superconductivity.\cite{rf:Laughlin} 
Experimentally, the pseudogap has actually been detected by probing the density of states (DOS) directly. 
The angle-resolved photoemission spectroscopy (ARPES)\cite{rf:Ding,rf:Loeser1,rf:Loeser2} and the scanning tunneling spectroscopy\cite{rf:Renner,rf:Kugler} measurements have shown that the pseudogap in the normal state develops continuously into the superconducting gap at low temperatures below the superconducting transition temperature $T_{\rm c}$.
On the other hand, it has been pointed out from the tunneling conductance measurements that the pseudogap is distinct even below $T_{\rm c}$.\cite{rf:Krasrov,rf:Suzuki,rf:Yurgens}
The experimental verification of the pseudogap has not been settled yet.

The pseudogap should be suppressed by the application of a magnetic field if it relates to the superconductivity. 
Therefore, the investigation of magnetic-field response is important for clarifying the mechanism of the pseudogap formation. 
From the NMR measurements in optimally doped YBa$_2$Cu$_3$O$_{7-\delta}$, the pseudogap has been claimed to be suppressed\cite{rf:Mitrovic} or be insensitive to magnetic fields.\cite{rf:Gorny}
In the underdoped YBa$_2$Cu$_4$O$_8$, no field effect on the onset temperature of the pseudogap has been reported,\cite{rf:Zheng1} while another NMR study has indicated a clear field effect in slightly overdoped TlSr$_2$CaCu$_2$O$_{6.8}$.\cite{rf:Zheng2}
Accordingly, current situation on the magnetic field effect of the pseudogap has also been controversial.

The out-of-plane resistivity $\rho_{\rm c}$ has been regarded as an effective probe for the pseudogap study, since $\rho_{\rm c}$ directly measures Cooper pair- or quasiparticle-tunnelings between CuO$_2$ layers in both normal and superconducting states,\cite{rf:Morozov,rf:Shibauchi} providing the DOS at the Fermi energy. 
Moreover, theoretical works have suggested that $\rho_{\rm c}$ reflects the magnitude of the momentum at ($\pi$, 0) points on the anisotropic Fermi surface.\cite{rf:Ioffe,rf:Valla} where the pseudogap first opens up.\cite{rf:Norman}
Thus, $\rho_{\rm c}$ should be sensitive to the pseudogap formation. 
In several high-$T_{\rm c}$ cuprates, actually, the depletion of DOS has been evidenced by a semiconductor-like upturn of $\rho_{\rm c}$ observed below a characteristic temperature $T^*$.\cite{rf:Watanabe,rf:Lavrov}
On the basis of these features, the magnetic field dependence of the pseudogap has been tried to be probed by the $c$-axis magnetoresistance (MR) measurement. 
The MR measurement in Bi$_2$Sr$_{2-x}$La$_x$CuO$_{6+\delta}$ by Lavrov {\it et al.}\cite{rf:Lavrov} has shown that the MR in heavily underdoped nonsuperconducting samples is quite small below $T^*$, while the noticeable negative MR appears in superconducting samples. 
The results have implied two different kinds of the pseudogap in cuprates. 
That is, one is insensitive to the magnetic field, while another is sensitive. 
Then, the observed negative MR in the superconducting sample may indicate the suppression of the pseudogap correlated with the superconductivity. 
In order to establish the relation to the superconductivity clearly, more systematic investigations in other systems must be needed.

In this paper, we report the temperature dependence of $\rho_{\rm c}$ in systematically oxygen-controlled single crystals Bi$_{1.74}$Pb$_{0.38}$Sr$_{1.88}$CuO$_{6+\delta}$ in magnetic fields up to 15 T in order to investigate the relation between the pseudogap formation and the superconductivity. 
The characteristic temperatures defined by the MR results are discussed from the point of the pseudogap formation. 
Finally the phase diagram obtained in the present study is compared to other Bi-based high-$T_{\rm c}$ cuprates.

Single crystals of Bi$_{1.74}$Pb$_{0.38}$Sr$_{1.88}$CuO$_{6+\delta}$ were grown by the floating-zone method in a similar way as reported by Chong {\it et al}.\cite{rf:Chong}
The crystals were then characterized using x-ray back-Laue photography and confirmed having single phase by means of powder x-ray diffraction method.
The typical size of a single crystal measured was 1.5 $\times$ 1.0 $\times$ 0.05 mm$^3$. 
The hole concentration was controlled by annealing under vacuum, flowing Ar, flowing O$_2$ or 7 atm O$_2$ atmosphere. 
$T_c$ of samples was determened by measuring the magnetic susceptibility with a SQUID magnetometer (Quantum Design, MPMS-XL5) in a magnetic field of 1 Oe. 
Table \ref{tbl:1} shows the achieved $T_{\rm c}$ after annealing in the listed conditions. 
We have succeeded in controlling the hole concentration from heavily underdoped nonsuperconducting region to heavily overdoped nonsuperconducting region. 
Here, $T_{\rm c} =$ 0 K means that a sample does not indicate any superconducting behaviors down to 1.5 K.
The electrical resistivity parallel to the $c$-axis was measured by a standard DC four-terminal method. 
Magnetic fields were applied up to 15 T parallel to the $c$-axis by a superconducting magnet.

\begin{figure}[p]
\begin{center}
\includegraphics[width=0.4\linewidth]{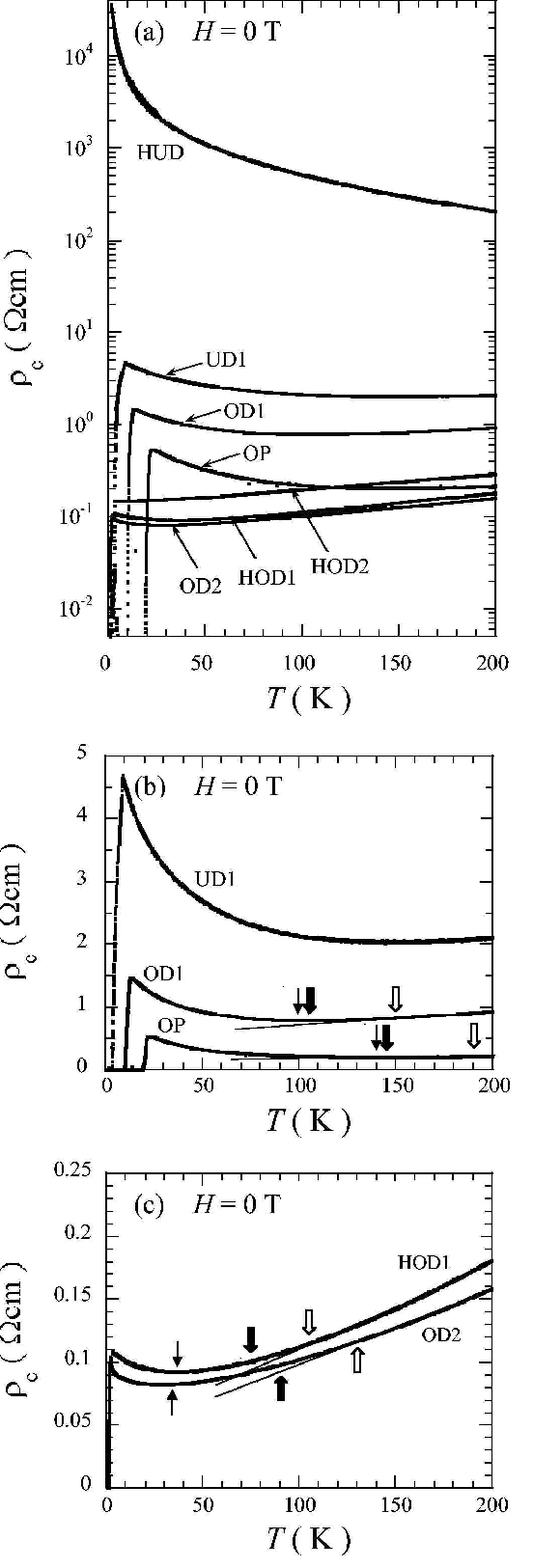}
\end{center}
\caption{Temperature dependence of the out-of-plane resistivity $\rho_{\rm c}$ of Bi$_{1.74}$Pb$_{0.38}$Sr$_{1.88}$CuO$_{6+\delta}$ in zero field for (a) all samples, (b) UD1, OP and OD1 and (c) OD2 and HOD1. 
Open thick, filled thick and thin arrows denote temperatures $T^*_{0.01}$, $T^*_{0.1}$ and $T^*_{\rm min}$ where $\rho_{\rm c}$ deviates from its linear extrapolation by 1 \% and 10 \% and exhibits the local minimum, respectively. 
}\label{fig:1}
\end{figure}

Figure \ref{fig:1}(a) shows the temperature dependence of the out-of-plane resistivity $\rho_{\rm c}$ in Bi$_{1.74}$Pb$_{0.38}$Sr$_{1.88}$CuO$_{6+\delta}$ with different oxygen concentrations listed in Table \ref{tbl:1}. 
In heavily underdoped nonsuperconducting sample HUD, $\rho_{\rm c}$ exhibits a semiconductor-like temperature dependence at low temperatures below 200 K. 
Through the hole-doping, $\rho_{\rm c}$ decreases and represents a superconducting transition in UD1, OP, OD1, and OD2. 
In such superconducting samples, $\rho_{\rm c}$ exhibits a metallic temperature dependence followed by a semiconductor-like upturn at low temperatures. 
A crossover temperature $T^*$ between the metallic and semiconductor-like $\rho_{\rm c}$ is defined here by using three different criteria; 10 \% and 1 \% deviation of $\rho_{\rm c}$ at $T^*_{0.1}$ and $T^*_{0.01}$ from the linear extrapolation of $\rho_{\rm c}$ from higher temperature, and the local minimum of $\rho_{\rm c}$ at $T^*_{\rm min}$. 
Open thick, filled thick and thin arrows in Figs. \ref{fig:1}(b) and \ref{fig:1}(c) indicate $T^*_{0.01}$, $T^*_{0.1}$ and $T^*_{\rm min}$, respectively. 
Among these criteria, $T^*_{0.1}$ has been used to define the pseudogap temperature by Lavrov {\it et al.}.\cite{rf:Lavrov} 
In heavily overdoped nonsuperconducting samples, the upturn becomes weak and the $\rho_{\rm c}$ value decreases simultaneously with increasing hole concentration. 
The increase of the $\rho_{\rm c}$ value may be induced by the randomness due to the excess oxygen. 
In the most heavily overdoping sample HOD2, the upturn of $\rho_{\rm c}$ disappears. 

One may consider that these overall features of the hole-doping dependence of $\rho_{\rm c}$ and $T_{\rm c}$ are similar to those of other systems such as Bi$_{2}$Sr$_{2}$CaCu$_2$O$_{8+\delta}$\cite{rf:Watanabe} and Bi$_{2}$Sr$_{2-x}$La$_x$CuO$_{6+\delta}$.\cite{rf:Lavrov} 
The present systematic doping experiment reveals that the semiconductor-like $\rho_{\rm c}$ at low temperatures appears even in the heavily overdoped nonsuperconducting sample HOD1. 
The present result strongly indicates that the pseudogap formed below $T^*$ exists in the heavily overdoped nonsuperconducting sample as well as the superconducting one, although the close correlation between $T^*$ and the superconductivity has been claimed so far.\cite{rf:Timusk,rf:Shibauchi}

\begin{figure}[p]
\begin{center}
\includegraphics[width=0.7\linewidth]{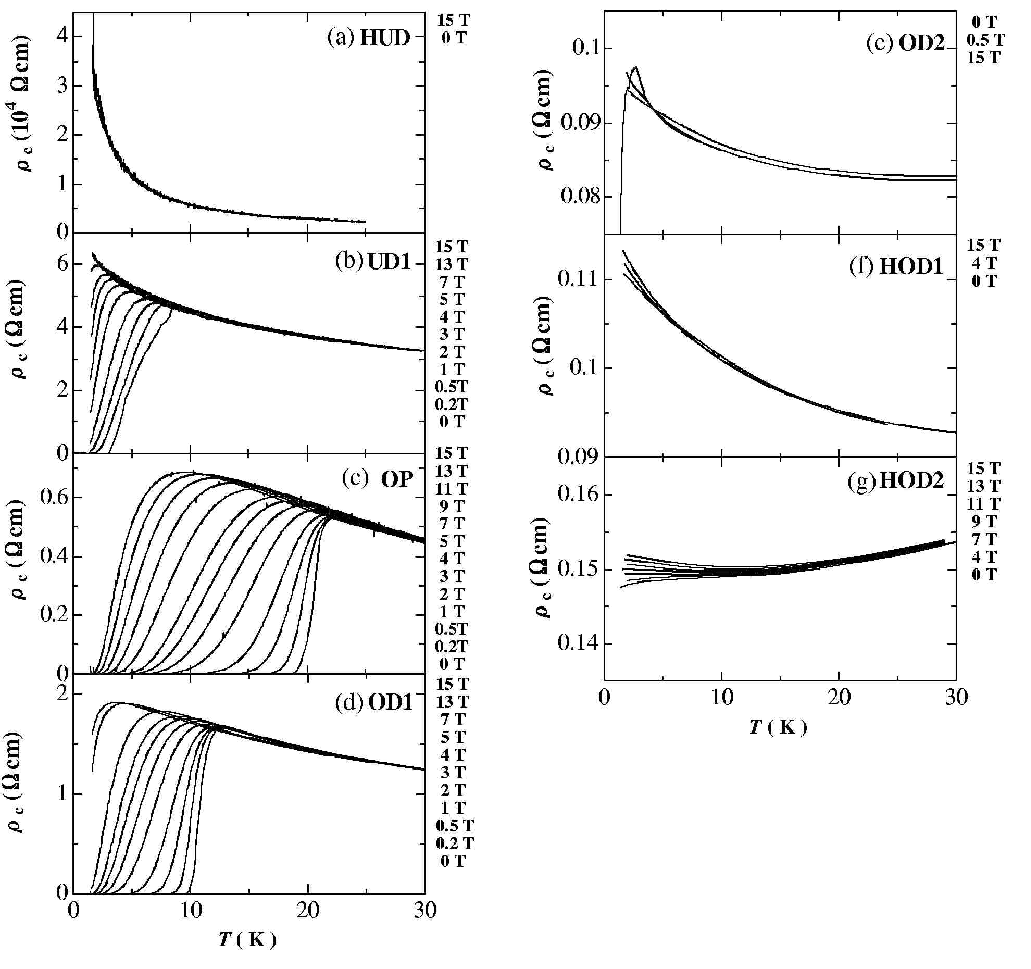}
\end{center}
\caption{Temperature dependence of the out-of-plane resistivity $\rho_{\rm c}$ of Bi$_{1.74}$Pb$_{0.38}$Sr$_{1.88}$CuO$_{6+\delta}$ in magnetic fields parallel to the $c$-axis. 
}\label{fig:2}
\end{figure}

Figure \ref{fig:2} shows the temperature dependence of $\rho_{\rm c}$ in magnetic fields parallel to the $c$-axis. 
In HUD, as is similarly reported in Bi$_{2}$Sr$_{2-x}$La$_x$CuO$_{6+\delta}$,\cite{rf:Lavrov} the upturn of $\rho_{\rm c}$ is slightly enhanced in a magnetic field. 
Similar positive MR is found in HOD1 and HOD2 (Figs. \ref{fig:2}(f) and \ref{fig:2}(g)). 
On the other hand, the upturn is suppressed by the application of magnetic fields in UD1, OP, OD1 and OD2(Figs. \ref{fig:2}(b)-\ref{fig:2}(e)), which exhibit superconductivity. 
Such a suppression of the upturn of MR can be understood in a scenario of a pseudogap suppressed by a magnetic field.
Lavrov {\it et al.}\cite{rf:Lavrov} have concluded that such magnetic field sensitive pseudogap appears only in the superconducting samples.

\begin{figure}[p]
\begin{center}
\includegraphics[width=0.6\linewidth]{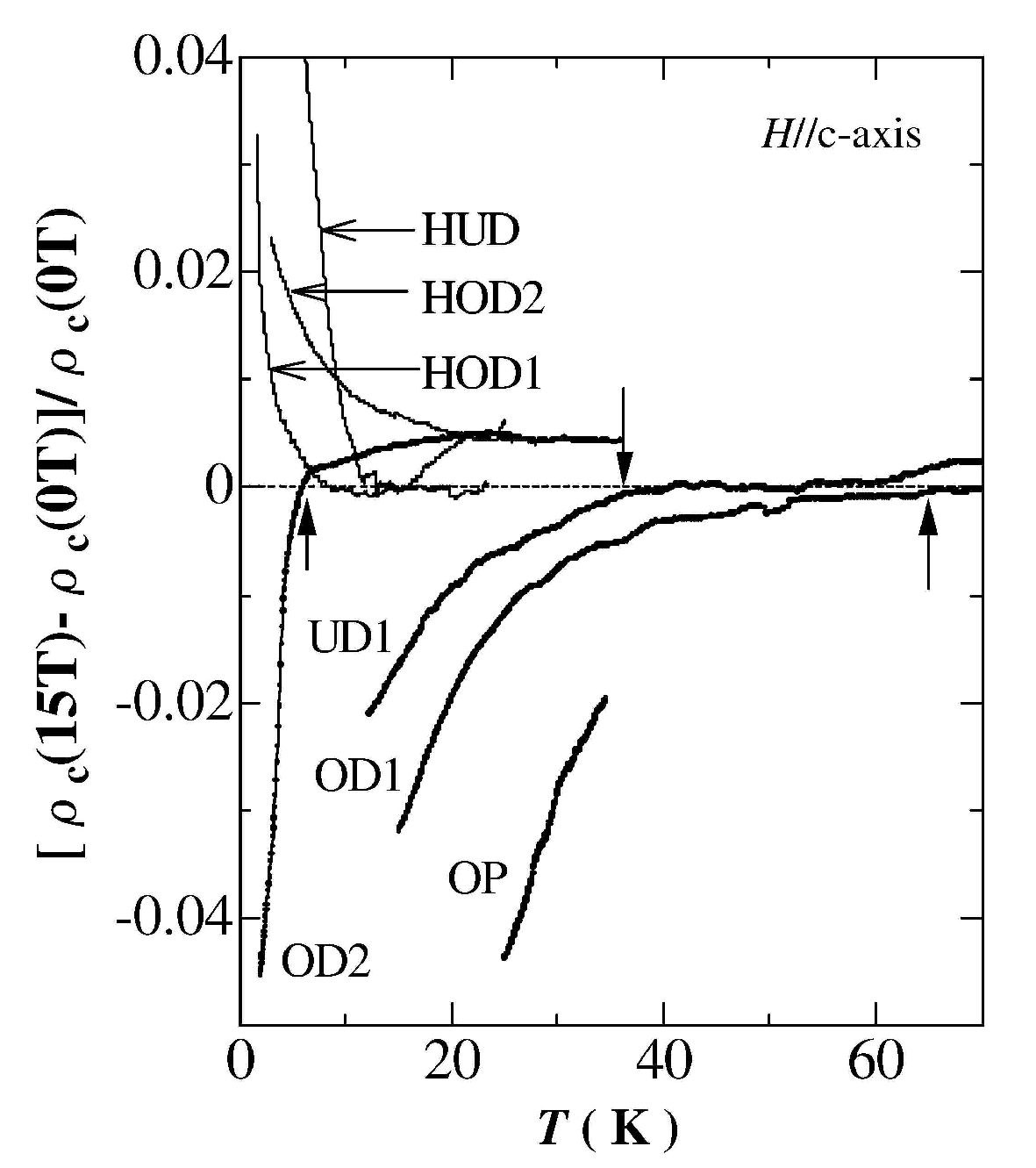}
\end{center}
\caption{Temperature dependence of the magnetoresistance of $\rho_{\rm c}$ of Bi$_{1.74}$Pb$_{0.38}$Sr$_{1.88}$CuO$_{6+\delta}$ in a magnetic field of 15 T parallel to the $c$-axis. 
Arrows denote the temperature $T^{**}$ where the negative magnetoresistance reaches 0.1 \%. 
}\label{fig:3}
\end{figure}

Figure \ref{fig:3} shows the temperature dependence of the MR of $\rho_{\rm c}$ in a magnetic field of 15 T parallel to the $c$-axis. 
Corresponding to the suppression of the upturn by the application of magnetic fields, the MR in superconducting samples becomes negative at low temperatures below $T^{**}$ which is defined as the temperature where the negative MR reach 0.1 \%, as indicated by arrows in Fig. \ref{fig:3}. 
Meanwhile, the positive MR is observed and $T^{**}$ cannot be defined in HUD and HOD1, though the upturn of $\rho_{\rm c}$ appears at low temperatures below $T^*$. 
The different magnetic-field dependence of MR below $T^*$ and $T^{**}$ suggests two different origins of the semiconductor-like upturn of $\rho_{\rm c}$, that is, two different kinds of pseudogaps formed at $T^*$ and $T^{**}$.

\begin{figure}[p]
\begin{center}
\includegraphics[width=0.7\linewidth]{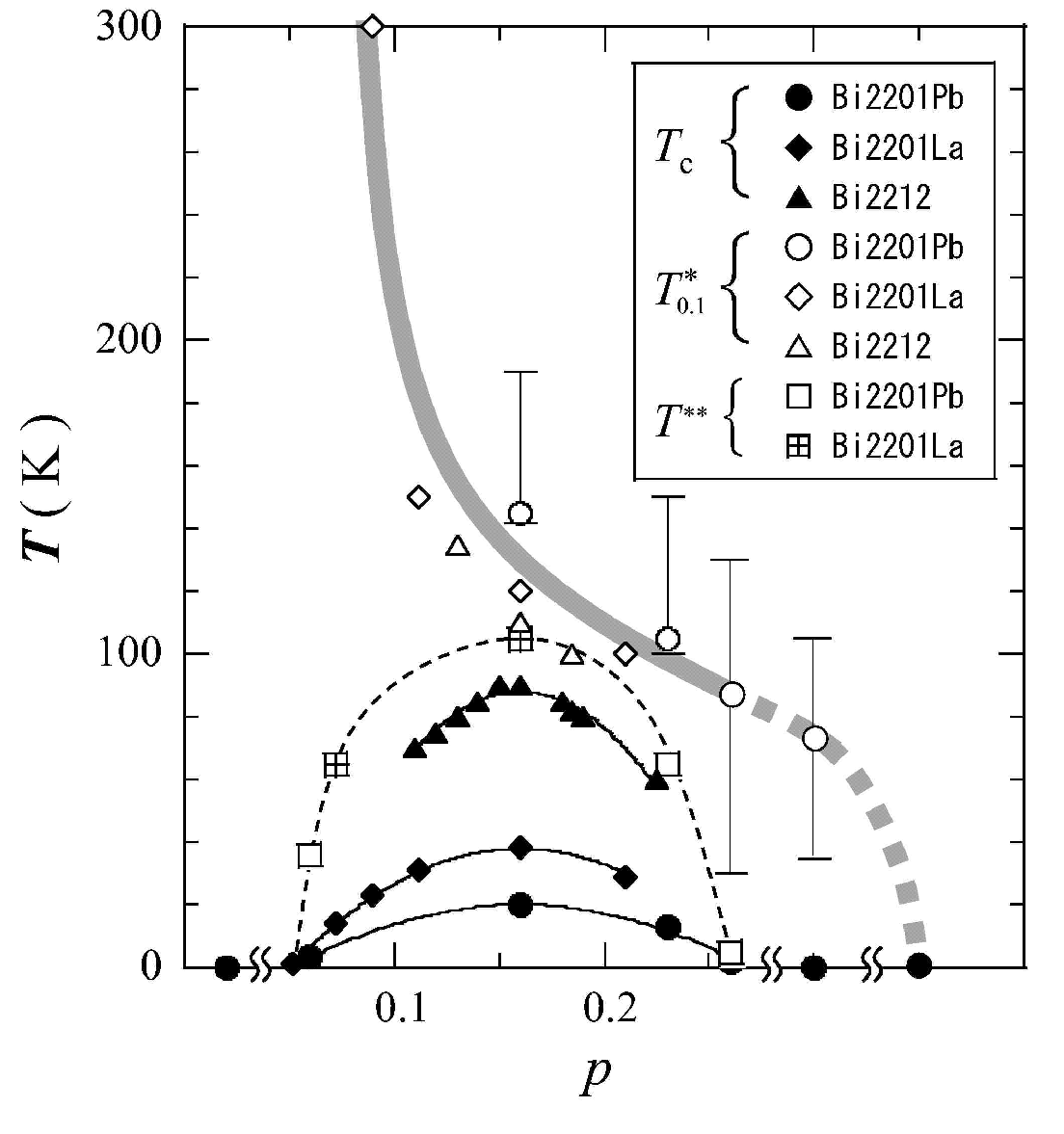}
\end{center}
\caption{Hole-concentration dependence of $T_{\rm c}$, $T^*$ and $T^{**}$. 
Bi2201Pb, Bi2201La and Bi2212 mean Bi$_{1.74}$Pb$_{0.38}$Sr$_{1.88}$CuO$_{6+\delta}$, Bi$_{2}$Sr$_{2-x}$La$_{x}$CuO$_{6+\delta}$\cite{rf:Lavrov} and Bi$_2$Sr$_2$CaCu$_2$O$_{8+\delta}$,\cite{rf:Watanabe} respectively. 
Vertical bars above and below $T^*_{0.1}$ ($\circ$) represent $T^*_{0.01}$ and $T_{\rm min}$ of Bi2201Pb, respectively. 
Curves are guides for eyes. 
}\label{fig:4}
\end{figure}

In order to discuss the relation between the pseudogap and the superconductivity, characteristic temperatures $T^*$, $T^{**}$ and $T_{\rm c}$ are plotted in the $T - p$ plane as shown in Fig. \ref{fig:4}. 
Hole-concentration $p$ of the superconducting samples is determined by using the empirical law proposed by Presland {\it et al}.\cite{rf:emp} 
The nonsuperconducting samples, however, are assumed to be plotted at relatively reasonable positions for $p$. 
For comparison, $T^*_{0.1}$, $T^{**}$ and $T_{\rm c}$ of Bi$_2$Sr$_2$CaCu$_2$O$_{8+\delta}$\cite{rf:Watanabe,rf:determine} and Bi$_{2}$Sr$_{2-x}$La$_{x}$CuO$_{6+\delta}$\cite{rf:Lavrov} are also plotted. 
Solid, broken and thick curves corresponding to $T_{\rm c}$, $T^*$, and $T^{**}$, respectively, are guides for eyes. 
$T^*$ of Bi$_{1.74}$Pb$_{0.38}$Sr$_{1.88}$CuO$_{6+\delta}$ decreases with increasing hole concentration. 
It is noted that $T^*(p)$ behavior is universal among such Bi-based cuprate superconductors,\cite{rf:Watanabe,rf:Lavrov} although the corresponding $T_{\rm c}$ are quite different in each system. 
Moreover, the result suggests that the pseudogap formed below $T^*$ exists even in the nonsuperconducting samples.
Owing to these features, it can be concluded that the correlation between $T^*$ and $T_{\rm c}$ is very weak. 
$T^*$ may relate to the different energy scale excitation such as the antiferromagnetic excitation due to the exchange interaction between Cu$^{2+}$ spins. 
In fact, $T^*$-curve resembles the $p$ dependence of the effective exchange interaction energy obtatined from the two magnon peak in the Raman scattering experiments.\cite{rf:Sugai}

In contrast to the $p$ dependence of $T^*$, $T^{**}$ in both Bi$_{1.74}$Pb$_{0.38}$Sr$_{1.88}$CuO$_{6+\delta}$ and Bi$_{2}$Sr$_{2-x}$La$_{x}$CuO$_{6+\delta}$ exhibit the same bell-shaped $p$ dependence shown by the broken curve in Fig. \ref{fig:4}. 
Such two different kinds of universal $p$ dependence of $T^*$ and $T^{**}$ in Bi-based cuprate superconductors demonstrate that the nonsuperconducting heavily overdoped samples possess only one pseudogap formed below $T^*$. 
The result means also that the superconductivity appears only below $T^{**}$, indicating that the pseudogap opened at $T^{**}$ must closely relate to the superconductivity.

A precursor of the superconductivity could develop at low temperatures below $T^{**}$. 
Actual $T_{\rm c}$, however, is much smaller than $T^{**}$ and those have large system dependence. 
Recently, Eisaki {\it et al.}\cite{rf:Eisaki} have proposed that $T_{\rm c}$ increases with decreasing the disorder in the A-site, which corresponds to the Sr-site in Bi-based high-$T_{\rm c}$ cuprates. This is consistent with the smallest $T_{\rm c}$ of Bi$_{1.74}$Pb$_{0.38}$Sr$_{1.88}$CuO$_{6+\delta}$ in Fig. \ref{fig:4}, because such a disorder becomes largest in the present system among Bi-based high-$T_{\rm c}$ cuprates on account of the nonstoichiometric Sr content and/or the partial substitution of Bi for Sr.

The present phase diagram is somewhat different in the $p$ dependence of $T^*$ from generally proposed two types. 
One has a quantum critical hole concentration $p_{\rm cr} \sim$ 0.19 at $T = 0$ and $T^*$ crosses $T_{\rm c}$, pointing to $p_{\rm cr}$ at $T =$ 0 K. 
Another shows that $T^*$ merges with $T_{\rm c}$ in the overdoped region.\cite{rf:Timusk,rf:Tallon,rf:Uchida} 
The difference could be attributed to the relative variation of $T_{\rm c}$ in comparison to the universal $T^*$ curve.
In Fig. \ref{fig:4}, if the $T_{\rm c}^{\rm max}$ were as large as $T^*$, the $T^*$-curve would cross the bell shape of $T_{\rm c}$ around the optimal doping. 
In this case, $T^*$ looks to fall to zero at $p_{\rm cr}$ or merges with $T_{\rm c}$ as proposed so far. 
In the present system, much smaller $T_{\rm c}^{\rm max}$ than $T^*$ results in the observed wide pseudogap region up to nonsuperconducting heavy doping. 
Assuming the linear hole concentration dependence of $T^*$ in heavily overdoped region, $T^*$ in the present system disappears at $p_{\rm cr} \sim$ 0.38, implying that the quantum critical point does not relate to the superconductivity in the present system.

The high magnetic field experiments were partly supported by the High Field Laboratory for Superconducting Materials, IMR, Tohoku University. 
This work was supported by a Grant-in Aid for Scientific Research from JSPS. 


\begin{references}
\bibitem{rf:Timusk} For a review, T. Timusk and B. Statt, Rep. Prog. Phys. {\bf 62}, 61 (1999).
\bibitem{rf:Randeria} M. Randeria, N. Trivedi, A. Moreo and R. T. Scalettar, Phys. Rev. Lett. {\bf 69}, 2001 (1992).
\bibitem{rf:Emery} V. J. Emery and S. A. Kivelson, Nature (London) {\bf 374}, 434 (1995). 
\bibitem{rf:Kao} Ying-Jer Kao, A. P. Iyengar, Q. Chen and K. Levin, Phys. Rev. B {\bf 64}, 140505(R) (2001). 
\bibitem{rf:Iyenger} A. P. Iyenger, Ying-Jer Kao, Q. Chen, K. Levin, J. Phys. Chem. Solids {\bf 63}, 2349 (2002). 
\bibitem{rf:Laughlin} S. Chakravarty, R. B. Laughlin, D. K. Morr and C. Nayak, Phys. Rev. B {\bf 63}, 094503 (2001).
\bibitem{rf:Ding} H. Ding, T. Yokoya, J. C. Campuzano, T. Takahashi, M. Randeria, M. R. Norman, T. Mochiku, K. Kadowaki and J. Giapintzakis, Nature (London) {\bf 382}, 51 (1996).
\bibitem{rf:Loeser1} A. G. Loeser, Z. -X. Shen, D. S. Dessau, D. S. Marshall, C. H. Park, P. Fournier, A. Kapitulnik, Science {\bf 273}, 325 (1996). 
\bibitem{rf:Loeser2} A. G. Loeser, Z. -X. Shen, M. C. Schabel, C. Kim, M. Zhang, A. Kapitulnik and P. Fournier, Phys. Rev. B {\bf 56}, 14185 (1997).
\bibitem{rf:Renner} Ch. Renner, B. Revaz, J. -Y. Genoud, K. Kadowaki and $\O$. Fischer, Phys. Rev. Lett. {\bf 80}, 149 (1998). 
\bibitem{rf:Kugler} M. Kugler, $\O$. Fischer, Ch. Renner, S. Ono and Y. Ando, Phys. Rev. Lett. {\bf 86}, 4911 (2001). 
\bibitem{rf:Krasrov} V. M. Krasnov, A. Yurgens, D. Winkler, P. Delsing and T. Claeson, Phys. Rev. Lett. {\bf 84}, 5860 (2000). 
\bibitem{rf:Suzuki} M. Suzuki and T. Watanabe, Phys. Rev. Lett. {\bf 85}, 4787 (2000). 
\bibitem{rf:Yurgens} A. Yurgens, D. Winkler, T. Claeson, S. Ono and Y. Ando, Phys. Rev. Lett. {\bf 90}, 147005 (2003). 
\bibitem{rf:Mitrovic} V. F. Mitrovi\'{c}, H. N. Bachman, W. P. Halperin, M. Eschrig, J. A. Sauls, A. P. Reyes, P. Kuhns and W. G. Moulton, Phys. Rev. Lett. {\bf 82}, 2784 (1999).
\bibitem{rf:Gorny} K. Gorny, O. M. Vyaselev, J. A. Martindale, V. A. Nandor, C. H. Pennington, P. C. Hammel, W. L. Hults, J. L. Smith, P. L. Kuhns, A. P. Reyes and W. G. Moulton, Phys. Rev. Lett. {\bf 82}, 177 (1999).
\bibitem{rf:Zheng1} G. -Q. Zheng, W. G. Clark, Y. Kitaoka, K. Asayama, Y. Kodama, P. Kuhns and W. G. Moulton, Phys. Rev. B {\bf 60}, R9947 (1999).
\bibitem{rf:Zheng2} G. -Q. Zheng, H. Ozaki, W. G. Clark, Y. Kitaoka, P. Kuhns, A. P. Reyes, W. G. Moulton, T. Kondo, Y. Shimakawa and Y. Kubo, Phys. Rev. Lett {\bf 85}, 405 (2000).
\bibitem{rf:Morozov} N. Morozov, L. Krusin-Elbaum, T. Shibauchi, L. N. Bulaevskii, M. P. Maley, Y. I. Latyshev and T. Yamashita, Phys. Rev. Lett. {\bf 84}, 1784 (2000).
\bibitem{rf:Shibauchi} T. Shibauchi, L. Krusin-Elbaum, M. Li, M. P. Maley and P. H. Kes, Phys. Rev. Lett. {\bf 86}, 5763 (2001).
\bibitem{rf:Ioffe} L. B. Ioffe and A. J. Millis, Phys. Rev. B {\bf 58}, 11631 (1998).
\bibitem{rf:Valla} T. Valla, A. V. Fedorov, P. D. Johnson, Q. Li, G. D. Gu and N. Koshizuka, Phys. Rev. Lett. {\bf 85}, 828 (2000).
\bibitem{rf:Norman} M. R. Norman, H. Ding, M. Randeria, J. C. Campuzano, T. Yokoya, T. Takeuchi, T. Takahashi, T. Mochiku, K. Kadowaki, P. Guptasarma and D. G. Hinks, Nature (London) {\bf 392}, 157 (1998).
\bibitem{rf:Watanabe} T. Watanabe, T. Fujii and A. Matsuda, Phys. Rev. Lett. {\bf 84}, 5848 (2000).
\bibitem{rf:Lavrov} A. N. Lavrov, Y. Ando and S. Ono, Europhys. Lett. {\bf 57}, 267 (2002).@
\bibitem{rf:Chong} I. Chong, T. Terashima, Y. Bando, M. Takano, Y. Matsuda, T. Nagaosa and K. Kumagai, Physica C {\bf 290}, 57 (1997).
\bibitem{rf:emp} The empirical relation between $T_{\rm c}/T_{\rm c}^{\rm max}$ and $p$, $T_{\rm c}/T_{\rm c}^{\rm max} = 1 - 82.6(p - 0.16)^2$, has been proposed by Presland {\it et al.}\cite{rf:Presland} on the basis of $T_{\rm c}$'s in Bi- and Tl-based high-$T_{\rm c}$ superconductors with various oxygen stoichiometry.
\bibitem{rf:Presland} M. R. Presland, J. L. Tallon, R. G. Buckley, R. S. Liu and N. E. Flower, Physica C {\bf 176}, 95 (1991).
\bibitem{rf:determine} For comparison, $T^*_{0.1}$ of Bi$_2$Sr$_2$CaCu$_2$O$_{8+\delta}$ is determined here by using the data in Ref. \onlinecite{rf:Watanabe} where $T^*$ has been defined as the temperature where $\rho_{\rm c}$ deviates from the linear extrapolation. 
\bibitem{rf:Sugai} S. Sugai, H. Suzuki, Y. Takayanagi, T. Hosokawa and N. Hayamizu, Phys. Rev. B {\bf 68}, 184504 (2003). 
\bibitem{rf:Eisaki} H. Eisaki, N. Kaneko, D. L. Feng, A. Damascelli, P. K. Mang, K. M. Shen, Z. -X. Shen and M. Greven, Phys. Rev. B {\bf 69}, 064512 (2004).
\bibitem{rf:Tallon} For a review, J. L. Tallon and J. W. Loram, Physica C {\bf 349}, 53 (2001).
\bibitem{rf:Uchida} S. Uchida, Solid State Commun. {\bf 126}, 57 (2003).
\end{references}


\begin{table}[h]
\caption[smallcaption]{The annealing condition and $T_{\rm c}$ of single crystals used in this study. 
$T_{\rm c} =$ 0 K means that a sample does not indicate any superconducting behaviors down to 1.5 K.}
\label{tbl:1}
\begin{ruledtabular}
\begin{tabular}{llllll}
   sample        & $T_{\rm c}$      & atmosphere &        temperature        &  period \\
                 &     ( K )        &            &        ( $^\circ$C )      &  ( day ) \\
\hline
   HUD        & 0          & vacuum         & 650   & 11   \\
   UD1        & 3.5        & vacuum         & 500   & 1   \\
   OP         & 20         & vacuum         & 650   & 4   \\
   OD1        & 13         & Ar    1 atm    & 650   & 2   \\
   as grown   & 7.0        & --             & --    & --  \\
   OD2        & 1.0\footnotemark[1]$^, $\footnotemark[2]         & O$_2$ 1 atm    & 500   & 7   \\
   HOD1       & 0\footnotemark[2]          & O$_2$ 1 atm    & 500   & 7   \\
   HOD2       & 0          & O$_2$ 7 atm    & 450   & 7    \\
\end{tabular}
\end{ruledtabular}
\footnotetext[1]{$T_{\rm c}$ was determined by the resistivity measurement down to 0.5K.}
\footnotetext[2]{OD2 and HOD1 are different batch samples.}
\end{table} 

\end{document}